\documentclass[conference]{IEEEtran}
\IEEEoverridecommandlockouts
\usepackage{cite}
\usepackage{amsmath,amssymb,amsfonts}
\usepackage{algorithmic}
\usepackage{graphicx}
\usepackage{textcomp}
\usepackage{xcolor}
\usepackage{bm}
\usepackage{latexsym}
\def\qed{\hfill $\Box$}
\newtheorem{theorem}{Theorem}
\newtheorem{prop}{Proposition}
\newtheorem{lemma}{Lemma}
\def\BibTeX{{\rm B\kern-.05em{\sc i\kern-.025em b}\kern-.08em
    T\kern-.1667em\lower.7ex\hbox{E}\kern-.125emX}}

\IEEEoverridecommandlockouts\IEEEpubid{\makebox[\columnwidth]{ 978-1-6654-3540-6/22~\copyright~2022 IEEE \hfill} \hspace{\columnsep}\makebox[\columnwidth]{ }}
\begin{document}

\title{MMSE Signal Detection for MIMO Systems \\
based on Ordinary Differential Equation
\thanks{This work was supported by JSPS Grant-in-Aid for Scientific Research(A)
Grant Number JP22H00514.}
}

\author{\IEEEauthorblockN{Ayano Nakai-Kasai}
\IEEEauthorblockA{\textit{Graduate School of Engineering} \\
\textit{Nagoya Institute of Technology}\\
Nagoya, Japan \\
nakai.ayano@nitech.ac.jp}
\and
\IEEEauthorblockN{Tadashi Wadayama}
\IEEEauthorblockA{\textit{Graduate School of Engineering} \\
\textit{Nagoya Institute of Technology}\\
Nagoya, Japan \\
wadayama@nitech.ac.jp}
}

\maketitle

\begin{abstract}
Motivated by emerging technologies for energy efficient analog computing 
and continuous-time processing, 
this paper proposes continuous-time minimum mean squared error estimation 
for multiple-input multiple-output (MIMO) systems 
based on an ordinary differential equation.
Mean squared error (MSE) is a principal detection performance measure 
of estimation methods for MIMO systems.
We derive 
an analytical MSE formula that indicates the MSE at any time.
The MSE of the proposed method depends on a regularization parameter 
which affects the convergence property of the MSE. 
Furthermore, 
we extend the proposed method by using a time-dependent regularization parameter 
to achieve better convergence performance.
Numerical experiments indicated excellent agreement with the theoretical values 
and improvement in the convergence performance owing to the use of the time-dependent parameter.
\end{abstract}

\begin{IEEEkeywords}
Ordinary differential equation, MIMO, MMSE estimation, analog computing
\end{IEEEkeywords}

\section{Introduction}
In the next generation wireless communication systems, beyond 5G and 6G, 
massive connectivity should be achieved with 
ultra high speed and large capacity communication \cite{6G}. 
The number of mobile devices increases every year, 
and the traffic and computational loads
at the base stations are becoming heavier.
It has been pointed out that there remain various implementation challenges 
with regard to the fulfillment of the demand for large-scale signal processing 
in base stations of the next generation wireless network systems \cite{KLi}.
In particular, 
typical signal detection methods in multiple-input multiple-output (MIMO) 
systems such as zero-forcing and minimum mean squared error (MMSE) \cite{Kim} 
detection methods depend on centralized processing at the base station and 
require a heavy computational burden for the matrix inversion computation,
which requires, in general, a cubic time complexity.
The significant amount of signal detection loads in a base station 
has become a major bottleneck in the implementation of the next generation systems \cite{6G}.
Massive parallel computation 
with matrix inversion hardware \cite{Chetan} may be one of the solutions 
but it needs tremendous energy consumption.
Therefore, there are strong demands to develop novel signal 
processing methods to achieve reasonable signal detection performance with
high energy efficiency. One possibility to ease the bottleneck
would be to reconsider analog-domain signal processing.

%The large-scale computation 
%becomes a serious problem in other research areas as well.
%Especially in studies of deep learning, 

%analog computing is gaining attention again 
%Estimations by deep learning usually require a large number of training data, 
%especially in recent applications such as networks for Internet of Things, voice/image recognition systems, and self-driving techniques.
%The increase of data, however, occupies a large portion of computing resources at the training process.
%This can be a serious problem 
%when the computing resource is not provided only for the training task 
%but assumed to perform multiple tasks.
%The situation is common 
%except in applications operated by a center specializing in the data processing.
%In such a case, the computational efficiency, 
%i.e., the number of operations per energy consumption per time, 
%becomes a main concern in the implementation of deep learning.
%It has been argued since the 1960s that 
%matrix operations such as multiplying and accumulation 
%can be operated in parallel at constant time by using analog calculations in memory \cite{Steinbuch}.

Recently, in the field of deep learning (DL), 
analog-domain computation has regained researchers' attention
mainly from the perspective of computational efficiency 
\cite{JWelser, Haensch, EACartier}.
Analog optical computing such as 
a photonic chip-based neural network (NN) 
proposed in \cite{Lin} also has several advantages 
such as high computational efficiency, scalability, and stability.
One of the recent studies was a complex-valued NN on a photonic chip 
proposed by Zhang et al. \cite{HZhang}.
In addition to this, the optical computation is expected to play an 
important role 
in solving large-scale problems 
such as combinatorial optimizations or probabilistic graphical models \cite{Ab}.

These works inspire us to exploit such analog-domain signal processing 
not only for deep neural networks but also for 
wireless communication networks.
An analog computer is fundamentally composed of
analog adders, multipliers, integrators, and 
other nonlinear devices, 
and it can simulate any linear/nonlinear 
ordinary differential equations (ODEs).
If one can formulate a high-dimensional 
signal detection task as a continuous-time 
dynamical system, it can be implemented with analog devices,
and we can expect that they will provide high energy efficiency.
%Therefore, it is worth studying possibility of analog-domain
%high-dimensional signal processing tasks.

Another advantage of continuous-time dynamical systems 
for a signal processing task is that they bring us an insight
into the discrete-time algorithms for solving the task,
which is a counterpart of the continuous system.
Neural ODE\cite{RTQChen} is an ODE including a NN, 
i.e, its dynamics can be learned from data.
Any numerical method for solving neural ODEs such as 
the Euler method and the Runge-Kutta method 
can be used for discretizing a high-dimensional neural ODE.
The correspondence between the continuous neural dynamical system 
and the discrete-time inference procedure opens a new way to 
understand the property of the discrete-time procedure.

In this paper, we revisit analog-domain computing as a tool for 
overcoming the computing bottleneck at the base station in wireless communications and 
explore new signal detection methods.
We present a continuous-time MMSE signal detection method for MIMO systems, 
which is derived directly as a form of ODE without any matrix inversion computation.
With the benefit of the ODE representation, we can obtain theoretical analyses of 
the ODE-based MMSE detection method for MIMO signals.
% An analog-domain MMSE MIMO detection problem 
% may be the right problem to tackle for solving the 
% bottleneck at the base station.
To the best of our knowledge, 
there are no directly relevant proposals and analyses 
in previous literature.

%This enables us analyses of stability and convergence behavior as discussed in \cite{YZhang1}--\cite{LXiao}.
The main contributions of this paper are listed below:
\begin{enumerate}
    \item We propose a continuous-time MMSE detection method for MIMO signals.
            The method includes a regularization parameter that controls convergence behavior of the estimation method.
            We show the stability of the proposed method.
    \item We derive an analytical formula of mean squared error (MSE) in a closed form. 
            The MSE is the principal performance measure of signal detection methods. 
            The formula is based on the eigenvalue decomposition of the Gram matrix.
            From the MSE formula, we immediately derive the asymptotic MSE.
            These analyses enable us to track the quality of the estimation at any time instant.
    \item We introduce a time-dependent regularization parameter to achieve improved convergence performance. 
            % This can be regarded as the typical scheduling of learning rate used in deep learning.
            We also derive an analytical MSE formula for the time-dependent system. 
            Numerical experiments will show that 
            the convergence performance is improved 
            by optimizing the time-dependent regularization parameter.
\end{enumerate}

Analog computing for high-dimensional signal processing 
is a developing technology from a hardware point of view, 
but we believe that the analysis of the proposed method 
is a meaningful step toward 
forthcoming analog-domain high dimensional signal processing in wireless communications.

\section{Preliminaries}
\subsection{Notations}
In the rest of the paper, we use the following notations.
The superscript $(\cdot)^{\mathrm{H}}$ denotes the Hermitian transpose.
The zero vector and identity matrix are represented as $\bm{0}$ and $\bm{I}$, respectively.
$\ell_2$-norm is $\|\cdot\|$.
The complex circularly symmetric Gaussian distribution $\mathcal{CN}(\bm{0},\bm{\Sigma})$ 
has mean vector $\bm{0}$ and covariance matrix $\bm{\Sigma}$.
The expectation and trace operators are $\mathbb{E}[\cdot]$ and $\mathrm{Tr}[\cdot]$, respectively.
The diagonal matrix is given by $\mathrm{diag}[\ldots]$ with the diagonal elements shown in the square brackets.

\subsection{System Model}
\label{sec:system}
In this paper, we consider the following received signal model:
\begin{equation}
    \bm{y} = \bm{Hs} + \bm{w},
    \label{eq:y}
\end{equation}
where $\bm{y}\in\mathbb{C}^m$ is the received signal, 
$\bm{H}\in\mathbb{C}^{m\times n}$ is the channel matrix, 
$\bm{s}\in\mathbb{C}^n$ is the transmitted signal that follows $\mathcal{CN}(\bm{0},\bm{I})$, 
and $\bm{w}\in\mathbb{C}^m$ is the measurement noise that follows $\mathcal{CN}(\bm{0},\sigma^2\bm{I}$).
% In the following, a Gram matrix of the channel matrix, $\bm{H}^\mathrm{H}\bm{H}$, 
% is assumed to be full-rank 
% {\color{red}{but all the following derivations hold without the assumption}}.
In the following, the channel matrix $\bm{H}$ 
is assumed not to be a zero matrix.

A linear estimate $\hat{\bm{s}}:=\bm{Wy}$ for MIMO systems is characterized by the matrix $\bm{W}\in\mathbb{C}^{n\times m}$, 
which is determined according to each estimation method.
The matrix $\bm{W}$ for MMSE signal detection \cite{Kim} can be obtained by minimizing the MSE 
given by $\mathbb{E}[\|\bm{Wy}-\bm{s}\|^2]$.
The resulting MMSE estimate is derived as 
\begin{equation}
    \hat{\bm{s}} = \left(\bm{H}^\mathrm{H}\bm{H}+\sigma^2\bm{I}\right)^{-1}\bm{H}^\mathrm{H}\bm{y}.
    \label{eq:mmse}
\end{equation}

\section{MMSE Estimation as ODE}
In the case of continuous-time systems, 
it is often difficult to calculate the inverse of a matrix \cite{YZhang4}, 
so that deriving the MMSE estimate \eqref{eq:mmse} is not straightforward.
This paper considers a gradient dynamical system for the MMSE estimation 
and describes the evolution of the estimate in continuous-time systems.

A function 
\begin{equation}
    f(\bm{x}) := \|\bm{y}-\bm{Hx}\|^2 + \eta\|\bm{x}\|^2, 
    \label{eq:cost}
\end{equation}
where $\eta>0$, 
can be regarded as the objective function for MMSE signal detection 
because the unique stationary point of $f(\bm{x})$ coincides with 
the MMSE estimate \eqref{eq:mmse} when $\eta=\sigma^2$ \cite{Li}.
The scalar value $\eta$ in \eqref{eq:cost} behaves as a regularization parameter.
The gradient vector of $f(\bm{x})$ is given by 
$\nabla f(\bm{x})=(\bm{H}^\mathrm{H}\bm{H}+\eta\bm{I})\bm{x}-\bm{H}^\mathrm{H}\bm{y}$.

In this paper, 
we regard the objective function \eqref{eq:cost} as a potential function of 
a continuous-time gradient dynamical system.
We then obtain an estimate, $\bm{x}(t)$, of the transmitted signal $\bm{s}$ 
at time $t\geq 0$ that evolves 
according to the ODE 
\begin{equation}
    \frac{d\bm{x}(t)}{dt} =-\nabla f(\bm{x}(t)) 
    = -(\bm{H}^\mathrm{H}\bm{H}+\eta\bm{I})\bm{x}(t)+\bm{H}^\mathrm{H}\bm{y}.
    \label{eq:ode}
\end{equation}
We further assume the initial condition 
$\bm{x}(0) = \bm{H}^\mathrm{H}\bm{y}$.
We name the proposed signal detection based on the ODE \eqref{eq:ode} 
Ordinary Differential Equation-based MMSE (ODE-MMSE) method.

A closed-form representation of the estimate $\bm{x}(t)$ can be derived 
by the solution for a first-order linear ODE with constant coefficients \cite{ode}.
This gives an analytical insight into the ODE-MMSE method discussed in the next section.
\begin{prop}
    The estimate of ODE-MMSE method at time $t\geq0$ that follows the ODE \eqref{eq:ode} is 
    represented as a random vector depending on the transmitted signal vector $\bm{s}$ 
    and the noise vector $\bm{w}$, 
    and given by 
    \begin{align}
        \bm{x}(t) = (\bm{Q}(t)+\bm{R})(\bm{Hs}+\bm{w}), 
        \label{eq:solution}
    \end{align}
    where 
    \[\bm{Q}(t) := \exp{(-(\bm{H}^\mathrm{H}\bm{H}+\eta\bm{I})t)}
    \left(\bm{I}-(\bm{H}^\mathrm{H}\bm{H}+\eta\bm{I})^{-1}\right)\bm{H}^\mathrm{H}\]
    and 
    $\bm{R}:=(\bm{H}^\mathrm{H}\bm{H}+\eta\bm{I})^{-1}\bm{H}^\mathrm{H}$.
\end{prop}
\textit{Proof}: 
An equilibrium point $\bm{x}^\ast$ of the ODE \eqref{eq:ode} can be obtained 
as the solution of the equation $\frac{d\bm{x}(t)}{dt}=0$.
This is given by $\bm{x}^\ast=(\bm{H}^\mathrm{H}\bm{H}+\eta\bm{I})^{-1}\bm{H}^\mathrm{H}\bm{y}$.
The equilibrium point is unique 
because the potential function \eqref{eq:cost} is strictly convex.
We define the residual error vector between $\bm{x}(t)$ and the equilibrium point as 
$\bm{e}(t):=\bm{x}(t)-\bm{x}^\ast$, 
and then the ODE \eqref{eq:ode} can be replaced with 
\begin{align}
    \frac{d\bm{e}(t)}{dt} = \frac{d\bm{x}(t)}{dt} 
    = -(\bm{H}^\mathrm{H}\bm{H}+\eta\bm{I})\bm{e}(t).
    \label{eq:et}
\end{align}
This is the typical first-order linear ODE with constant coefficients and 
can be solved with a matrix exponential \cite{ode}.
The solution is given by 
\begin{align}
    \bm{e}(t) &= \exp\left(-(\bm{H}^\mathrm{H}\bm{H}+\eta\bm{I})t\right)\bm{e}(0) \nonumber \\
    &= \exp\left(-(\bm{H}^\mathrm{H}\bm{H}+\eta\bm{I})t\right)\left(\bm{I}-(\bm{H}^\mathrm{H}\bm{H}+\eta\bm{I})^{-1}\right)\bm{H}^\mathrm{H} \bm{y}.
    \label{eq:eteq}
\end{align}
Therefore, the solution of \eqref{eq:ode} can be obtained 
by substituting \eqref{eq:eteq} and \eqref{eq:y} for $\bm{x}(t) = \bm{e}(t) + \bm{x}^\ast$, 
and by summarizing the terms of the equation.
\qed

% Equation \eqref{eq:solution} shows the evolution of the MMSE estimate in continuous-time systems.

The stability of the system \eqref{eq:et} can be evaluated 
via the eigenvalues of the matrix $\bm{A}:=\bm{H}^\mathrm{H}\bm{H}+\eta\bm{I}$.
\begin{prop}
    \label{prop:stability}
    The system \eqref{eq:et} is stable.
\end{prop}
\textit{Proof}: 
From \eqref{eq:et}, the stability of the system depends on the Hermitian matrix $-\bm{A}=-(\bm{H}^\mathrm{H}\bm{H}+\eta\bm{I})$.
The Hermitian matrix $\bm{H}^\mathrm{H}\bm{H}$ is positive semidefinite 
and the matrix $\eta\bm{I}$ is positive definite.
The Hermitian matrix $-\bm{A}$ becomes negative definite 
so that it only has real and negative eigenvalues.
From these facts, the system \eqref{eq:et} is proven to be stable.
\qed

From Proposition~\ref{prop:stability}, ODE-MMSE method has the following property.
\begin{prop}
    ODE-MMSE method minimizes the objective function \eqref{eq:cost}.
\end{prop}
\textit{Proof}: 
The equilibrium point $\bm{x}^\ast$ is the unique point for minimizing the objective function \eqref{eq:cost} 
where the derivative equals zero.
From Proposition~\ref{prop:stability}, 
the estimate of the ODE-MMSE method is guaranteed to converge to the equilibrium point, i.e., the minimum value.
Therefore, the estimate of ODE-MMSE method converges to the unique point for minimizing the objective function.
\qed

The ODE \eqref{eq:ode} has a close relation to the complex-valued NN \cite{HZhang}.
This NN can be regarded as a signal detection system for MIMO 
by using the transmitted and received signals as the outputs and inputs of the NN, respectively.
Moreover, the elementwise equation of \eqref{eq:ode} has the same formulation as 
an output of the complex-valued NN which is represented 
by weighted sum of the complex inputs and bias.
This relation motivates the realization of the proposed ODE-MMSE method 
as well as the complex-valued NN.

\section{MSE Analysis}
In this section, we derive an analytical MSE formula 
and then verify the validity and the convergence property of the ODE-MMSE method 
by computer simulation.

\subsection{Derivation of Analytical MSE}
\label{sec:derivation}
The MSE between the estimate $\bm{x}(t)$ and the transmitted signal $\bm{s}$, 
\begin{equation}
    \mathrm{MSE}(t) := \mathbb{E}[\|\bm{x}(t)-\bm{s}\|^2], 
    \label{eq:originalmse}
\end{equation} 
is a principal performance indicator of MIMO signal detection methods \cite{Joham} 
but the analytical formula cannot always be derived.
For instance, in a signal detection method based on approximate message passing, 
the MSE is analyzed under the assumption of large system limit \cite{Hayakawa}.
However, 
the proposed method has the advantage that 
the analytical MSE formula can be described by a closed-form 
without any constraint on system parameters, 
which is shown in Theorem~\ref{theo:analyticalmse} below.

In this section, we derive an analytical MSE formula by using eigenvalue decomposition of 
the Gram matrix $\bm{H}^\mathrm{H}\bm{H}$.
Suppose that the Gram matrix is decomposed as 
$\bm{H}^\mathrm{H}\bm{H} = \bm{U}\mathrm{diag}[\lambda_1,\ldots,\lambda_n]\bm{U}^\mathrm{H}$,
where $\bm{U}\in\mathbb{C}^{m\times m}$ is a unitary matrix 
and $\lambda_1,\ldots,\lambda_n$ are nonnegative eigenvalues.
We assume $\lambda_1\geq\ldots\geq\lambda_n\geq0$ for convenience of the subsequent analyses.
By using the decomposition, the following theorem holds.
\begin{theorem}
    \label{theo:analyticalmse}
    The analytical MSE for ODE-MMSE method is given by 
    \begin{align}
        \mathrm{MSE}(t) &= \sum_{i=1}^n \frac{\lambda_i(\lambda_i+\eta-1)^2(\lambda_i+\sigma^2)e^{-2(\lambda_i+\eta)t}}{(\lambda_i+\eta)^2} \nonumber \\
        &\ - \sum_{i=1}^n \frac{2\lambda_i(\lambda_i+\eta-1)(\eta-\sigma^2)e^{-(\lambda_i+\eta)t}}{(\lambda_i+\eta)^2} \nonumber \\
        &\ + \sum_{i=1}^n \frac{\eta^2+\sigma^2\lambda_i}{(\lambda_i+\eta)^2}.
        \label{eq:mse}
    \end{align}
\end{theorem}
\textit{Proof}: 
Substituting \eqref{eq:solution} for 
the right-hand side of \eqref{eq:originalmse} yields 
\begin{align}
	\mathrm{MSE}(t)
	&= \mathbb{E}\left[\|\left((\bm{Q}(t) + \bm{R})\bm{H}-\bm{I}\right)\bm{s} 
    + \left(\bm{Q}(t) + \bm{R}\right)\bm{w}\|^2 \right] \nonumber \\
	% &= \mathrm{Tr}\left[\mathbb{E}[\bm{ss}^\mathrm{H}]((\bm{Q}(t)+\bm{R})\bm{H}-\bm{I})^\mathrm{H}((\bm{Q}(t)+\bm{R})\bm{H}-\bm{I})\right] \nonumber \\
	% &\ + \mathrm{Tr}\left[\mathbb{E}[\bm{ww}^\mathrm{H}](\bm{Q}(t)+\bm{R})^\mathrm{H}(\bm{Q}(t)+\bm{R})\right] \nonumber \\
	&= \mathrm{Tr}\left[((\bm{Q}(t)+\bm{R})\bm{H}-\bm{I})^\mathrm{H}((\bm{Q}(t)+\bm{R})\bm{H}-\bm{I})\right] \nonumber \\
	&\ + \sigma^2\mathrm{Tr}\left[(\bm{Q}(t)+\bm{R})^\mathrm{H}(\bm{Q}(t)+\bm{R})\right].
    \label{eq:tmpmse}
\end{align}
The matrix exponential $e^{-(\bm{H}^\mathrm{H}\bm{H}+\eta\bm{I})t}$ in $\bm{Q}(t)$
can be diagonalized by using the eigenvalues of the Gram matrix as 
\begin{equation}
	e^{-(\bm{H}^\mathrm{H}\bm{H}+\eta\bm{I})t} = \bm{U}\mathrm{diag}[e^{-(\lambda_1+\eta)t}, \ldots, e^{-(\lambda_n+\eta)t}]\bm{U}^\mathrm{H}.
\end{equation}
From the fact, the terms in \eqref{eq:tmpmse} can be diagonalized and calculated 
% by using the eigenvalues $\lambda_1,\ldots,\lambda_n$ and the parameter $\eta$.
as 
\begin{align}
	&\mathrm{Tr}[(\bm{Q}(t)+\bm{R})^\mathrm{H}(\bm{Q}(t)+\bm{R})] \nonumber \\
	&= \sum_{i=1}^n \frac{\lambda_i\left((\lambda_i+\eta-1)e^{-(\lambda_i+\eta)t}+1\right)^2}{(\lambda_i+\eta)^2} 
\end{align}
and 
\begin{align}
	&\mathrm{Tr}\left[\left((\bm{Q}(t)+\bm{R})\bm{H}-\bm{I}\right)^\mathrm{H}\left((\bm{Q}(t)+\bm{R})\bm{H}-\bm{I}\right)\right] \nonumber \\
	&= \sum_{i=1}^n \frac{\left(\lambda_i(\lambda_i+\eta-1)e^{-(\lambda_i+\eta)t}-\eta\right)^2}{(\lambda_i+\eta)^2},
\end{align}
respectively.
The detailed calculation is shown in Appendix.
The analytical formula \eqref{eq:mse} is obtained by summarizing the terms of the matrix exponential.
\qed

Theorem~\ref{theo:analyticalmse} explicitly gives the analytical MSE value 
of ODE-MMSE method at any time $t\geq0$.
By using this formula, we can describe an asymptotic MSE value, 
i.e., $\mathrm{MSE}(t)$ at the asymptotic limit of $t$.
Before that, we mention the MSE of MMSE estimation \eqref{eq:mmse} 
to derive the asymptotic MSE value.
\begin{lemma}
    The MSE of MMSE estimation \eqref{eq:mmse}, 
    $\mathrm{MSE}_\mathrm{mmse}:=\mathbb{E}[\|\hat{\bm{s}}-\bm{s}\|^2]$, is given by 
    \begin{equation}
        \mathrm{MSE}_\mathrm{mmse}=\sum_{i=1}^n \frac{\sigma^2}{\lambda_i+\sigma^2}.
        \label{eq:mseofmmse}
    \end{equation}
\end{lemma}
\textit{Proof}: 
This can be derived by using MMSE estimate \eqref{eq:mmse} and 
the eigenvalue decomposition of the Gram matrix. 
\qed
\begin{lemma}
    \label{lemma:asymptotic}
    The asymptotic MSE value for ODE-MMSE method, 
    $\mathrm{MSE}_\infty:=\lim_{t\to\infty}\mathrm{MSE}(t)$, is given by 
    \begin{equation}
        \mathrm{MSE}_\infty = \sum_{i=1}^n \frac{\eta^2+\sigma^2\lambda_i}{(\lambda_i+\eta)^2}.
        \label{eq:asymptoticmse}
    \end{equation}
    The inequality $\mathrm{MSE}_\mathrm{mmse}\leq\mathrm{MSE}_\infty$ holds and 
    the equality holds if and only if $\eta=\sigma^2$.
\end{lemma}
\textit{Proof}: 
In the case of $t\to\infty$, 
the first and second terms of \eqref{eq:mse} vanish 
because $\lambda_i\geq0$ for $i=1,\ldots,n$ and $\eta>0$.
The remaining term is the asymptotic MSE value.
The latter statement is supported by  
the difference between \eqref{eq:asymptoticmse} and \eqref{eq:mseofmmse} 
\[
    \mathrm{MSE}_\infty-\mathrm{MSE}_\mathrm{mmse}
    = \sum_{i=1}^n\frac{\lambda_i(\eta-\sigma^2)^2}{(\lambda_i+\eta)^2(\lambda_i+\sigma^2)}
\]
is always nonnegative and equals $0$ if and only if $\eta=\sigma^2$.
\qed

From Theorem~\ref{theo:analyticalmse} and Lemma~\ref{lemma:asymptotic}, 
we can find that the regularization parameter $\eta$ controls 
the convergence rate and the asymptotic MSE value of the ODE-MMSE method.
The convergence rate largely depends on behavior of the exponential terms in \eqref{eq:mse}.
The larger $\eta$ accelerates the decrease in the exponential terms, 
but the asymptotic MSE value could be large depending on the value of $\eta$.

\subsection{Numerical Examples}
\label{sec:sim1}
We show numerical examples to confirm validity of the analytical MSE formula \eqref{eq:mse} 
and to evaluate the influence of the parameter $\eta$ on the convergence rate and the asymptotic MSE value \eqref{eq:asymptoticmse}.

\begin{figure}[tbp]
    \centerline{\includegraphics[width=0.95\columnwidth]{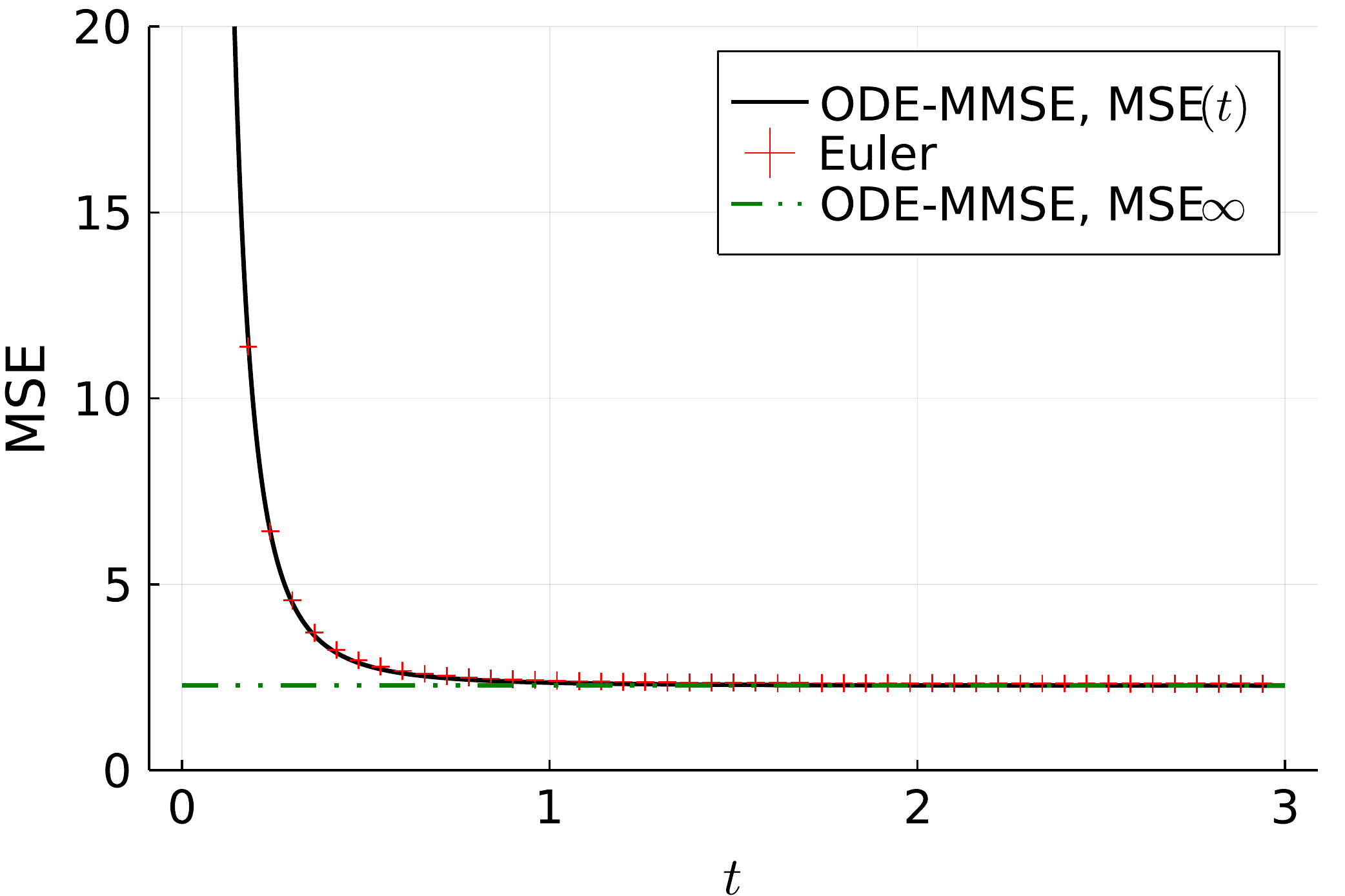}}
    \caption{Comparison of analytical MSE of ODE-MMSE with the arithmetic MSE of the Euler method, $(n,m,\sigma^2,\eta)=(8,8,1,0.5)$.}
    \label{fig:Euler}
\end{figure}

First, we evaluated the validity of the analytical MSE formula \eqref{eq:mse} 
by comparing with the arithmetic MSE obtained by Monte Carlo simulation under the single realization of the channel matrix $\bm{H}$.
Each element of the channel matrix $\bm{H}$ was generated by independent and identical distribution $\mathcal{CN}(0,1)$.
The system parameters were set to $(n,m,\sigma^2,\eta)=(8,8,1,0.5)$.
The horizontal line indicates the asymptotic MSE \eqref{eq:asymptoticmse}.
We employed the Euler method, where the behavior of $\bm{x}(t)$ can be determined directly by the ODE \eqref{eq:ode} 
and the estimate at time $t_N=\delta N$, where $\delta$ is step-size and $N=1,2,\ldots$, 
is given by 
\begin{equation}
    \bm{x}_{N} = \bm{x}_{N-1} -\delta(\bm{H}^\mathrm{H}\bm{H}+\eta\bm{I})\bm{x}_{N-1} + \delta\bm{H}^\mathrm{H}\bm{y}.
    \label{eq:euler}
\end{equation}
We set $\delta=10^{-3}$.
For the Monte Carlo simulation, 
pairs of $(\bm{s},\bm{w})$ and the corresponding received signal $\bm{y}$ 
were generated $1000$ times 
and the arithmetic MSE was computed.
Fig.~\ref{fig:Euler} shows the analytical MSE values of ODE-MMSE method, the arithmetic MSE values of the Euler method, 
and the asymptotic MSE value of ODE-MMSE method.
The curve of the analytical MSE formula is comparable to that of the Euler method 
with sufficient accuracy.
The analytical MSE value converges to the asymptotic MSE value.
The results strongly support the validity of the derivation of those theoretical results 
presented in Sect.~\ref{sec:derivation}.

\begin{figure}[tbp]
    \centerline{\includegraphics[width=0.95\columnwidth]{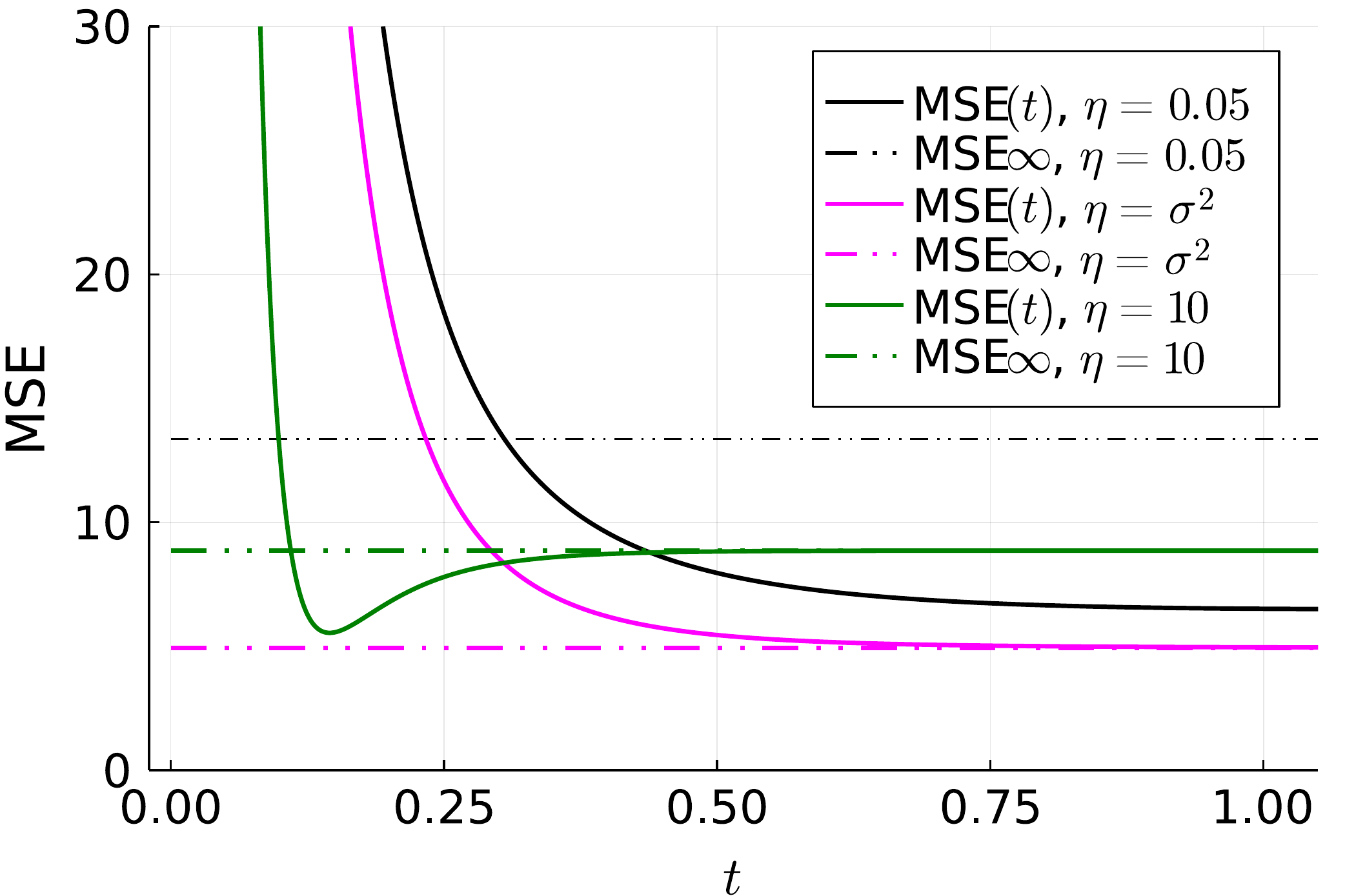}}
    \caption{Comparison of analytical MSE with different choices of the regularization parameter $\eta$, $(n,m,\sigma^2)=(32,32,1)$.}
    \label{fig:compareeta}
\end{figure}

Second, we evaluated the influence of the regularization parameter $\eta$ on the convergence behavior of ODE-MMSE method.
Specifically, we focused on the transitional behavior of ODE-MMSE method.
The system parameters were set to $(n,m,\sigma^2)=(32,32,1)$.
Fig.~\ref{fig:compareeta} shows the analytical MSE values with $\eta=0.05,\sigma^2$, and $10$.
The MSE with $\eta=10$ rapidly decreases 
but increases in the middle and finally show the higher asymptotic MSE value at the steady-state.
The result is consistent with the interpretation of \eqref{eq:mse} 
where larger $\eta$ accelerates decay of the exponential terms.
On the other hand, the decrease of the MSE with $\eta=0.05$ is the slowest but 
the MSE is lower in $0.5<t<1$ than that of $\eta=10$.
The asymptotic MSE is, however, the highest.
From these results, 
the convergence behavior largely depends on the choice of the regularization parameter $\eta$ 
and the superiority and inferiority of the MSE values can be switched depending on the time of interest.

\section{Time-dependent Control of Regularization Parameter}
\label{sec:odetime}
This section introduces time-dependent control of the regularization parameter 
aiming at the improvement of the convergence property of ODE-MMSE method.

\subsection{Derivation of Analytical MSE}
By the theoretical and simulation results in the previous section, 
we found that the regularization parameter $\eta$ significantly affects 
the convergence property of ODE-MMSE method.
Theorem~\ref{theo:analyticalmse} and Fig.~\ref{fig:compareeta} indicate that 
the larger $\eta$ yields faster convergence of ODE-MMSE method 
but yields the worse MSE value than the MMSE estimation ($\mathrm{MSE}_\infty$ with $\eta=\sigma^2$).
From these results, 
an adoption of time-dependent control of the regularization parameter $\eta$ is expected 
to hold both properties of faster convergence and the better asymptotic MSE value.
In this section, we improve ODE-MMSE method 
to be more flexible 
by employing the time-dependent regularization parameter $\eta(t)$. 

We consider an estimate of $\bm{s}$ 
that evolves according to the following ODE 
\begin{equation}
    \frac{d\bm{x}(t)}{dt} 
    = -(\bm{H}^\mathrm{H}\bm{H}+\eta(t)\bm{I})\bm{x}(t)+\bm{H}^\mathrm{H}\bm{y}.
    \label{eq:odetime}
\end{equation}
The expression $\eta(t)$ implies that 
the regularization parameter can vary depending on time $t$.
The initial condition is the same as that in \eqref{eq:ode}, i.e., 
$\bm{x}(0) = \bm{H}^\mathrm{H}\bm{y}$.
We name the proposed signal detection based on the ODE \eqref{eq:odetime} 
ODE-MMSE with time-dependent regularization parameter (tODE-MMSE) method.

The ODE \eqref{eq:odetime} can be solved by using variation of parameters method \cite{ode} 
because the matrix $\bm{A}(t):=\bm{H}^\mathrm{H}\bm{H}+\eta(t)\bm{I}$ is commutative.
\begin{prop}
    The estimate of tODE-MMSE method at time $t\geq0$ that follows the ODE \eqref{eq:odetime} 
    is given by 
    \begin{align}
        \bm{x}(t) &= \exp{\left(-\bm{H}^\mathrm{H}\bm{H}t-\xi(t) \bm{I}\right)} \nonumber \\
        &\ \times\left(\bm{I}+ \int_0^t e^{\bm{H}^\mathrm{H}\bm{H}u+\xi(u) \bm{I}}du\right) \bm{H}^\mathrm{H}\bm{y},  
        \label{eq:solutiontime}
    \end{align}
    where $\xi(T):=\int_0^T \eta(s)ds$.
\end{prop}
% \textit{Proof}: 
% The solution of the ODE \eqref{eq:odetime} is given by 
% the variation of parameters method as follows.
% \begin{align}
%     \bm{x}(t) &= e^{\int_0^t \bm{A}(s)ds}
%     \left(\bm{x}(0) + \int_0^t e^{-\int_0^u \bm{A}(s)ds}\bm{H}^\mathrm{H}\bm{y}du \right) \nonumber\\
%     &= e^{\int_0^t \bm{A}(s)ds}
%     \left(\bm{I} + \int_0^t e^{-\int_0^u \bm{A}(s)ds}du \right)\bm{H}^\mathrm{H}\bm{y} \nonumber.
% \end{align}
% \qed

Even in this case, an analytical MSE formula for \eqref{eq:solutiontime} can be derived 
in the same way as in Sect.~\ref{sec:derivation}.
\begin{theorem}
    The analytical MSE for the tODE-MMSE method is given by 
    \begin{align}
        \mathrm{MSE}(t) &= \sum_{i=1}^n \lambda_i(\lambda_i+\sigma^2)\!\left(1\!+\!\int_0^t e^{\lambda_iu+\xi(u)}du\right)^2 \!e^{-2(\lambda_i t+\xi(t))} \nonumber \\
        &\ -2\sum_{i=1}^n\lambda_i\left(1+\int_0^t e^{\lambda_iu+\xi(u)}du\right) e^{-(\lambda_i t+\xi(t))}+n.
        \label{eq:msetime}
    \end{align}
\end{theorem}
\textit{Proof}: 
MSE can be derived in the same procedure for Theorem~\ref{theo:analyticalmse} 
by using the eigenvalue decomposition of the Gram matrix.
Note that $\xi(t)$ is a scalar 
and that an integral in terms of a matrix is applied elementwise.
\qed

We can obtain the result of Theorem~\ref{theo:analyticalmse} 
by setting $\eta(t)=\eta$.
The analytical formula \eqref{eq:msetime} has a complicated form, 
but, like the case of the ODE-MMSE method, 
the form of time-dependent function $\eta(t)$ influences behavior of the estimation.

\subsection{Numerical Examples}
We show numerical examples to confirm validity of the analytical MSE formula \eqref{eq:msetime} 
and to compare the convergence performance of tODE-MMSE method 
with that of ODE-MMSE method.

The integral $\xi(t)=\int_0^t \eta(s)ds$ is analytically tractable in some cases.
For convenience, we use the following parametric model as the function $\eta(t)$, 
\begin{equation}
    \eta(t)=\frac{1}{\alpha t+\epsilon}+\sigma^2, 
    \label{eq:etainv}
\end{equation} 
where $\alpha$ is a parameter, and $\epsilon$ is a small number fixed to $10^{-8}$ in this paper.
The integral can be calculated as 
$\xi(t) = \frac{1}{\alpha}\log\left(\frac{\alpha t+\epsilon}{\epsilon}\right)+\sigma^2 t$.
The function $\eta(t)$ converges to $\sigma^2$ at the limit of $t\to\infty$.

\begin{figure}[tbp]
    \centerline{\includegraphics[width=0.95\columnwidth]{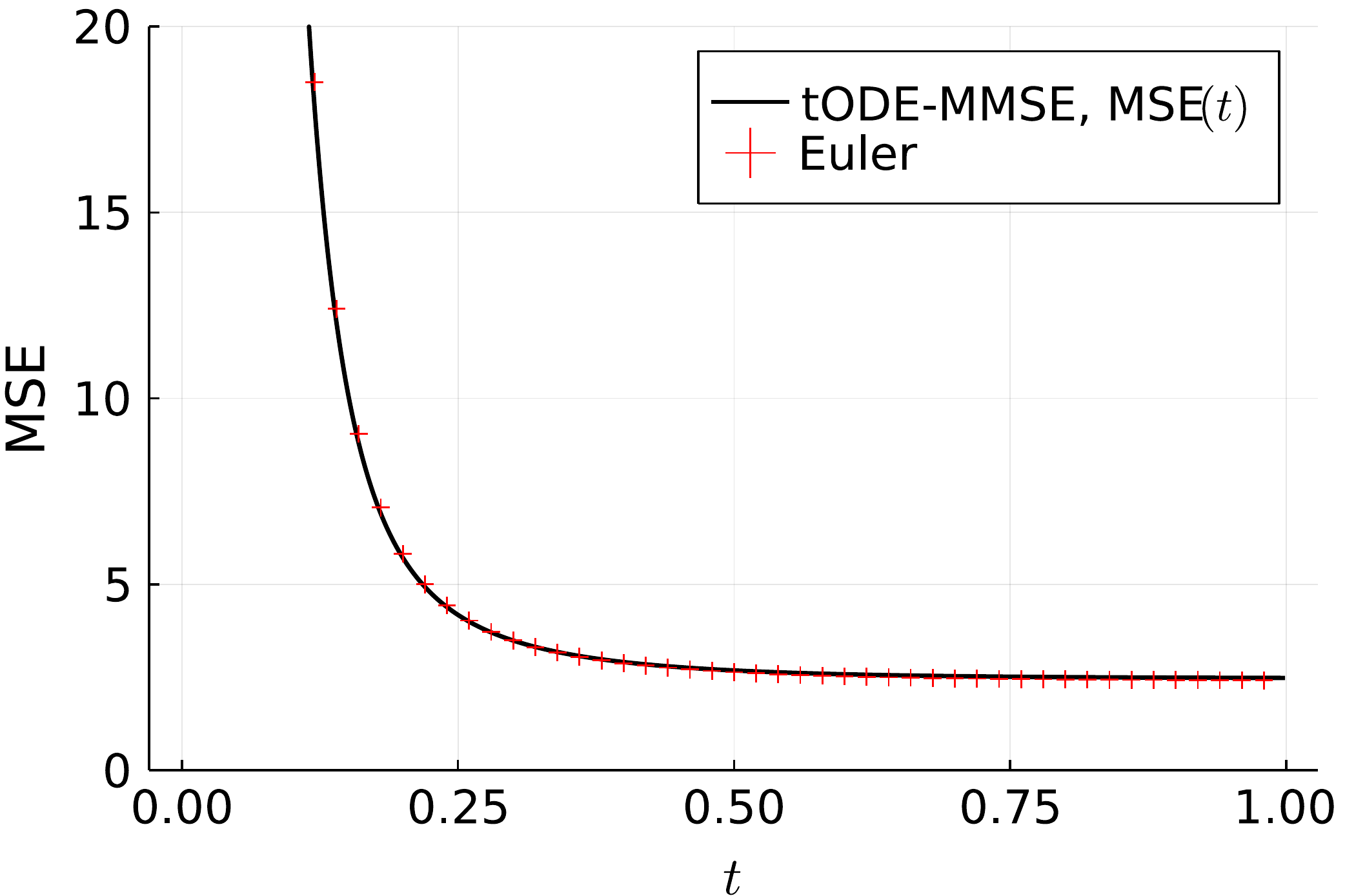}}
    \caption{Comparison of analytical MSE of tODE-MMSE with the arithmetic MSE of Euler method, $(n,m,\sigma^2)=(8,8,1)$.}
    \label{fig:Euleretat}
\end{figure}

We evaluated the validity of the analytical MSE formula \eqref{eq:msetime} 
by the comparison with the arithmetic MSE obtained by Monte Carlo simulation under the single realization of the channel matrix $\bm{H}$.
We employed the Euler method where $\eta$ in the equation \eqref{eq:euler} was replaced with $\eta(t_N)$.
The received and transmitted signals and the channel matrix were generated in the same way as in Sect.~\ref{sec:sim1}.
We used the tractable regularization function \eqref{eq:etainv} with $\alpha=500$.
The system parameters were set to $(n,m,\sigma^2)=(8,8,1)$.
Fig.~\ref{fig:Euleretat} shows the MSE values at time $t$ of the methods.
The curve of the analytical MSE formula is comparable to that of the Euler method, 
so that the validity of the analytical formula \eqref{eq:msetime} is supported.

Finally, we present an example of using the analytical MSE formula \eqref{eq:msetime} of tODE-MMSE method 
for improving the convergence property and 
compare the performance with that of ODE-MMSE method.
We have found in Fig.~\ref{fig:compareeta} that the performance of the proposed method 
largely depends on the choice of the regularization parameter. 
It is expected that 
we can improve the convergence property by tODE-MMSE method with an appropiate choice of the function $\eta(t)$.
There are various possible indicators to evaluate the goodness of convergence performance.
In this paper, we employed the functional 
\[F(\xi(t)) := \int_0^T \mathrm{MSE}(t) dt\] 
as the indicator.
If a method holds both properties of faster convergence and lower error, 
the value of the functional becomes smaller.
In the following, 
we optimize the parameter by minimizing the functional value.
Specifically, we choose the parameter that minimizes the functional by employing the grid search.
% We used QuadGK.jl \cite{quad} on Julia language \cite{julia}.

\begin{figure}[tbp]
    \centerline{\includegraphics[width=0.95\columnwidth]{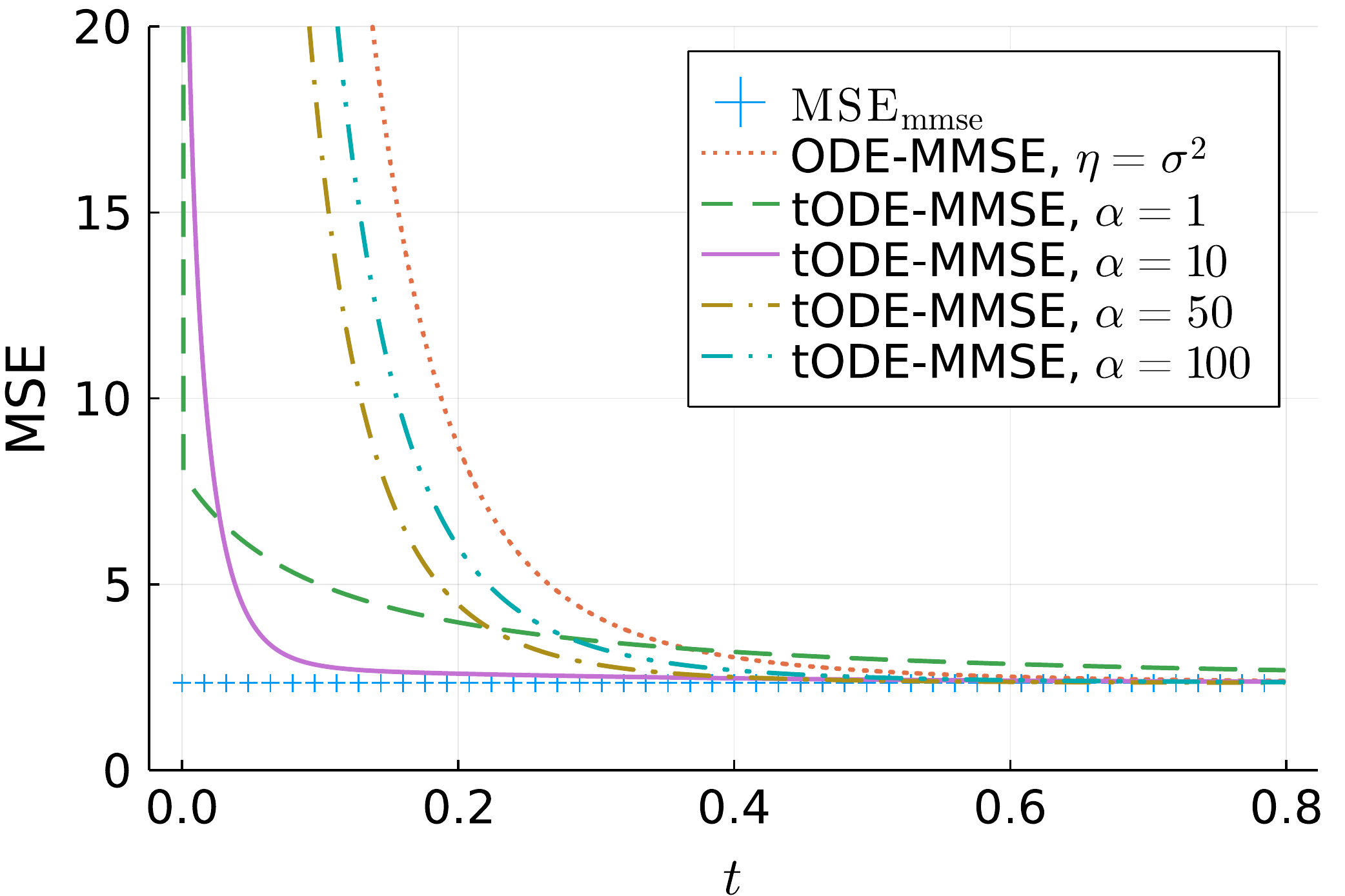}}
    \caption{The analytical MSE curves with different values of $\alpha$, $(n,m,\sigma^2)=(8,8,1)$.}
    \label{fig:gridsearch}
\end{figure}

We set $\alpha=1,10,50,100$ as the candidates for the parameter.
The system parameters were set to $(n,m,\sigma^2)=(8,8,1)$ and $T=0.8$.
Table~\ref{tab:tab1} summarizes the evaluated values of $F(\xi(t))=F(\alpha)$.
From the table, the value became the lowest with $\alpha=10$. 
% so that the convergence performance with $\alpha=10$ is expected to be the best.
Fig.~\ref{fig:gridsearch} shows the MSE of MMSE estimate $\mathrm{MSE}_\mathrm{mmse}$, 
the analytical MSE values of ODE-MMSE method with $\eta=\sigma^2$, 
and those of tODE-MMSE method with different values of $\alpha$.
From the figure, all the MSE curves of tODE-MMSE method converge to the value of $\mathrm{MSE}_\mathrm{mmse}$ 
faster than ODE-MMSE method.
Moreover, the convergence of the method with $\alpha=10$, 
which shows the lowest functional value in Table~\ref{tab:tab1}, 
is the fastest among the candidates.
This indicates that we can find an improved estimation method by the grid search using the functional value.
\begin{table}[tbp]
    \caption{Values of functional $F(\alpha)$.}
    \begin{center}
    \begin{tabular}{|c||c|c|c|c|c|}
        \hline
        $\alpha$ & $1$ & $10$ & $50$ & $100$ \\
        \hline
        $F(\alpha)$ & $2.9136$ & $2.4127$ & $12.4011$ & $17.5674$ \\
        \hline
    \end{tabular}
    \label{tab:tab1}
    \end{center}
\end{table}

\section{Conclusions}
Analog computing and continuous-time processing are gaining attention 
from the perspective of computational efficiency of DL and 
can be solutions to computational load problems in the next generation wireless communication systems.
Inspired by this background, 
we have considered continuous-time MMSE signal detection methods for MIMO systems.
We derived the continuous-time estimate as ODE and proposed ODE-MMSE method.
The analytical MSE formula is tractable 
by using eigenvalue decomposition of the Gram matrix of the channel matrix.
Simulation results showed the validity of the analytical MSE formula 
and the significant influence of the choice of the parameter $\eta$ on the convergence performance.
Moreover, we extended the ODE-MMSE method by introducing time-dependent parameter $\eta(t)$ and proposed tODE-MMSE method.
The validity of the analytical MSE formula for tODE-MMSE method and its convergence property were confirmed 
via computer simulation.

As a further development of this research, 
we can consider deriving a novel discrete-time algorithm 
by discretization of an evolution described by ODE. 
The approach in \cite{RTQChen} 
for obtaining output at any discrete time of NN from ODE can be applied to the development of signal detection algorithms.
The novel discrete-time signal detection algorithm is expected to be derived by considering ODE for signal detection.
% It is believed that a new discrete algorithm can be derived.

% \begin{table}[htbp]
% \caption{Table Type Styles}
% \begin{center}
% \begin{tabular}{|c|c|c|c|}
% \hline
% \textbf{Table}&\multicolumn{3}{|c|}{\textbf{Table Column Head}} \\
% \cline{2-4} 
% \textbf{Head} & \textbf{\textit{Table column subhead}}& \textbf{\textit{Subhead}}& \textbf{\textit{Subhead}} \\
% \hline
% copy& More table copy$^{\mathrm{a}}$& &  \\
% \hline
% \multicolumn{4}{l}{$^{\mathrm{a}}$Sample of a Table footnote.}
% \end{tabular}
% \label{tab1}
% \end{center}
% \end{table}

% \begin{figure}[htbp]
% \centerline{\includegraphics{fig1.png}}
% \caption{Example of a figure caption.}
% \label{fig}
% \end{figure}

\section*{Appendix}
\subsection{Derivation of Theorem~\ref{theo:analyticalmse}}
The matrix $\bm{B}(t):=\bm{Q}(t)+\bm{R}$ can be calculated as 
\begin{align}
    \bm{B}(t) &= \Bigl(\exp{(-(\bm{H}^\mathrm{H}\bm{H}+\eta\bm{I})t)}
    \left(\bm{I}-(\bm{H}^\mathrm{H}\bm{H}+\eta\bm{I})^{-1}\right) \nonumber \\
    & \ + (\bm{H}^\mathrm{H}\bm{H}+\eta\bm{I})^{-1}\Bigr)\bm{H}^\mathrm{H} \\
    &= \bm{U}\Bigl(\mathrm{diag}\Bigl[e^{-(\lambda_1+\eta)t}\left(1-\frac{1}{\lambda_1+\eta}\right), \ldots, \nonumber \\
    & \quad e^{-(\lambda_n+\eta)t}\left(1-\frac{1}{\lambda_n+\eta}\right)\Bigr] \nonumber \\
    & \ + \mathrm{diag}\left[\frac{1}{\lambda_1+\eta}, \ldots, \frac{1}{\lambda_n+\eta}\right]\Bigr)\bm{U}^\mathrm{H}\bm{H} \\
    &= \bm{U}\mathrm{diag}\Bigl[\frac{e^{-(\lambda_1+\eta)t}(\lambda_1+\eta-1)+1}{\lambda_1+\eta},\ldots, \nonumber \\
    & \quad \frac{e^{-(\lambda_n+\eta)t}(\lambda_n+\eta-1)+1}{\lambda_n+\eta}\Bigr]\bm{U}^\mathrm{H}\bm{H}
\end{align}
From \eqref{eq:tmpmse}, 
\begin{align}
    & \mathrm{Tr}\left[((\bm{Q}(t)+\bm{R})\bm{H}-\bm{I})^\mathrm{H}((\bm{Q}(t)+\bm{R})\bm{H}-\bm{I})\right] \nonumber \\
    &= \mathrm{Tr}\Biggl[\mathrm{diag}\Biggl[\left(\left(\frac{e^{-(\lambda_1+\eta)t}(\lambda_1+\eta-1)+1}{\lambda_1+\eta}\right)\lambda_1-1\right)^2,\ldots, \nonumber \\
    & \quad \left(\left(\frac{e^{-(\lambda_n+\eta)t}(\lambda_n+\eta-1)+1}{\lambda_n+\eta}\right)\lambda_n-1\right)\Biggr]\Biggr] \\
    &= \sum_{i=1}^n \frac{\left(\lambda_i(\lambda_i+\eta-1)e^{-(\lambda_i+\eta)t}-\eta\right)^2}{(\lambda_i+\eta)^2}
\end{align}
and 
\begin{align}
	&\mathrm{Tr}[(\bm{Q}(t)+\bm{R})^\mathrm{H}(\bm{Q}(t)+\bm{R})] \nonumber \\
	&= \mathrm{Tr}\Biggl[\mathrm{diag}\Bigl[\left(\frac{e^{-(\lambda_1+\eta)t}(\lambda_1+\eta-1)+1}{\lambda_1+\eta}\right)^2\lambda_1,\ldots, \nonumber \\
    & \quad \left(\frac{e^{-(\lambda_n+\eta)t}(\lambda_n+\eta-1)+1}{\lambda_n+\eta}\right)^2\lambda_n\Bigr]\Biggr] \\
    &= \sum_{i=1}^n \frac{\lambda_i\left((\lambda_i+\eta-1)e^{-(\lambda_i+\eta)t}+1\right)^2}{(\lambda_i+\eta)^2}.
\end{align}

\section{Derivation of Lemma~1}
\begin{align}
    &\mathrm{MSE}_\mathrm{mmse} \nonumber\\
    &=\mathbb{E}[\|\left(\bm{H}^\mathrm{H}\bm{H}+\sigma^2\bm{I}\right)^{-1}\bm{H}^\mathrm{H}(\bm{Hs}+\bm{w})-\bm{s}\|^2] \\
    &= \mathbb{E}\Biggl[\Bigl\|\left(\left(\bm{H}^\mathrm{H}\bm{H}+\sigma^2\bm{I}\right)^{-1}\bm{H}^\mathrm{H}\bm{H}-\bm{I}\right)\bm{s} \nonumber \\
    & \ +\left(\bm{H}^\mathrm{H}\bm{H}+\sigma^2\bm{I}\right)^{-1}\bm{H}^\mathrm{H}\bm{w}\Bigr\|^2\Biggr] \\
    &=\mathrm{Tr}\Bigl[\left(\left(\bm{H}^\mathrm{H}\bm{H}+\sigma^2\bm{I}\right)^{-1}\bm{H}^\mathrm{H}\bm{H}-\bm{I}\right)^\mathrm{H} \nonumber \\
    & \quad \cdot\left(\left(\bm{H}^\mathrm{H}\bm{H}+\sigma^2\bm{I}\right)^{-1}\bm{H}^\mathrm{H}\bm{H}-\bm{I}\right)\Bigr] \nonumber \\
    & \ + \sigma^2\mathrm{Tr}\Bigl[\left(\left(\bm{H}^\mathrm{H}\bm{H}+\sigma^2\bm{I}\right)^{-1}\bm{H}^\mathrm{H}\right)^\mathrm{H} \nonumber \\
    & \quad \cdot\left(\left(\bm{H}^\mathrm{H}\bm{H}+\sigma^2\bm{I}\right)^{-1}\bm{H}^\mathrm{H}\right)\Bigr] \\
    &= \sum_{i=1}^n \left(\frac{\lambda_i}{\lambda_i+\sigma^2}-1\right)^2 
    + \sigma^2\sum_{i=1}^n \left(\frac{\lambda_i}{(\lambda_i+\sigma^2)^2}\right) \\
    &= \sum_{i=1}^n \frac{\sigma^2}{\lambda_i+\sigma^2}
\end{align}

\section{Derivation of Theorem~2}
The matrix in \eqref{eq:solutiontime} can be decomposed as 
\begin{align}
    &\exp{\left(-\bm{H}^\mathrm{H}\bm{H}t-\xi(t) \bm{I}\right)}\left(\bm{I}+ \int_0^t e^{\bm{H}^\mathrm{H}\bm{H}u+\xi(u) \bm{I}}du\right) \nonumber \\
    &= \bm{U}\Bigl(\mathrm{diag}\Bigl[e^{-(\lambda_1 t+\xi(t))}\left(1+\int_0^t e^{\lambda_1 u+\xi(u)}du\right), \ldots, \nonumber \\
    & \quad e^{-(\lambda_n t+\xi(t))}\left(1+\int_0^t e^{\lambda_n u+\xi(u)}du\right)\Bigr]\Bigr)\bm{U}^\mathrm{H}.
\end{align}
By using this, the analytical MSE can be derived in the same way as Theorem~1.

\end{document}